\documentclass{article}
\usepackage{graphicx}
\usepackage{xcolor}
\usepackage[noadjust]{cite}
\author{
D. Micheli$^1$, R. Diamanti$^2$, L. Bastianelli$^{3,4}$, E. Colella$^4$, V. Mariani Primiani$^{3,4}$, \\
F. Moglie$^{3,4}$, A. Allasia$^5$, M. Crozzoli$^5$, M. Colombo$^6$ \\
  1 TIM S.p.A., Via Oriolo Romano, 240, Building B, 00189 Rome, Italy\\
  2 TIM S.p.A., Via Guido Miglioli 11, 60131 Ancona, Italy\\
  3 DII, Università Politecnica delle Marche, via Brecce Bianche 12, 60131, Ancona, Italy\\
  4 CNIT, Viale G.P. Usberti 181/A, 43124, Parma, Italy\\
  5 TIM S.p.A., Wireless Access Innovation – Telecom Italia, Turin, Italy\\
  6 Nokia Networks Italia, Energy Park 14, 20871 Vimercate, Italy\\

}

\begin{document}

\title{Test of 5G System in the Reverberation Chamber at mm-wave}

\maketitle

\begin{abstract}

The performance of a real fifth generation base station was studied by using a reverberation chamber as a
real life propagating environment.
Preliminary tests were conducted in order to define 5G base station operation conditions at mm-wave and emulated scenarios where reconfigurable intelligent surface(s) (RISs) will successively be tested.
Measurements campaign was carried out under the H2020 European project RISE-6G and a collaboration program between TIM S.p.A., Nokia and Universit\`a Politecnica delle Marche.

\end{abstract}

\vskip0.5\baselineskip

\section{Introduction}

5G technology, the fifth generation of the mobile telephony standard, represents the most advanced mobile data transmission technology currently available.
Although the name suggests a certain continuity with previous standards, in fact 5G is quite distinct from its predecessors.
5G, in fact, does not only represent the evolution of long-term 4G technology
(LTE) but allows a high speed of communication between a large number of low latency devices.
Mainly 5G is based on three evolutionary dimensions that have allowed this important generational step. The first important step achieved by the
technology was the increased speed of access to the network or enhanced Mobile Broad Band (eMBB) which allowed to
reach up to 10 Gbps. Increase in the number of connections to devices with low energy consumption denoted as massive
Machine Type Communications (mMTC) up to about 1 million devices per km$^2$. Also higher reliability and lower latency or
Ultra Reliable Low Latency Communications (URLLC) up to 1--2 ms. These characteristics, therefore, allow its application
in various sectors from autonomous vehicles, Internet of Things (IoT), remote controlling, machine-to-machine (M2M)
communication to smart environments capable of controlling and managing an enormous quantity of data. Compared to
previous technologies, an increase in mobile traffic of about 300~exabytes is expected to date by the end of 2027~\cite{ericsson}.
This is thanks to a wider spectrum frequency band that will cover millimeter waves (mmWave)~\cite{sakaguchi,Choudhury}. This spectrum guarantees a
high speed of mobile data transmission and a large number of interconnections, although the mmWave band shows an
excessive loss of path in the presence of obstacles. This problem can be overcome by means of intelligent antennas
equipped with high transmission power beamforming capable of overcoming multipath fading.
Recently, the emerging technology called reconfigurable intelligent surfaces (RISs) will be prominent for future wireless systems~\cite{rise1,rise2,rise3,rise4,rise5,rise6,rise7}.
The RIS integration with 5G systems and beyond is challenging in many aspects, e.g. from electromagnetic point-of-view, protocols, communication  and  localization, etc.
Results carried out in this measurements campaign will be preparatory for upcoming activities where RIS(s) will be tested.
This paper presents a preliminary experimental activity on the 5G base station (BS) functionalities, by carrying out over-the-air tests inside a reverberation chamber (RC)~\cite{micheli1,micheli2,micheli3,micheli4,kildal},
preparatory for future tests when the RIS will be included in the experimental set-up.
By using the RC we are able to reproduce complex propagation conditions both for the BS and UE.
Furthermore, the RC allows to perform tests in a controlled way without interferences coming from the external environment that could cause unwanted behavior at this stage.
Moreover, by opportunely load the chamber we are able to emulate a real-life environments, such as a commercial scenario, indoor/outdoor scenario and so on~\cite{dresda,remley,holloway}.
In order to replicate dynamic and/or static propagation conditions it is necessary to correctly set the RC by matching parameters of interest for wireless testing, i.e. the time delay spread, power delay profile.
The use of RC permits to test the BS and devices in various radio-channel conditions in the same laboratory.
This work is a collaboration between CNIT--Unit research of Ancona and TIM, both of them involved in the RISE-6G project, and Nokia.

\section{Experimental set-up and results}

Testing of wireless communication system and device in RC is nowadays considered very powerful from both mobile operator and BS manufacturing point of view~\cite{magazine}.
In fact, the RC is able to provide very rich multipath propagation environment, capable to stress the connection between the BS and UE(s)~\cite{micheli0,micheli0a,micheli1}.
A large variety of situations can be reproduced inside the RC~\cite{micheli1,micheli3,micheli4} by tuning RC parameters like the quality factor, the power delay profile and
the Rician K-factor by adding absorbing material inside it~\cite{dresda,chen,holloway2}.

\begin{figure}
\centering
\includegraphics[width=0.8\columnwidth]{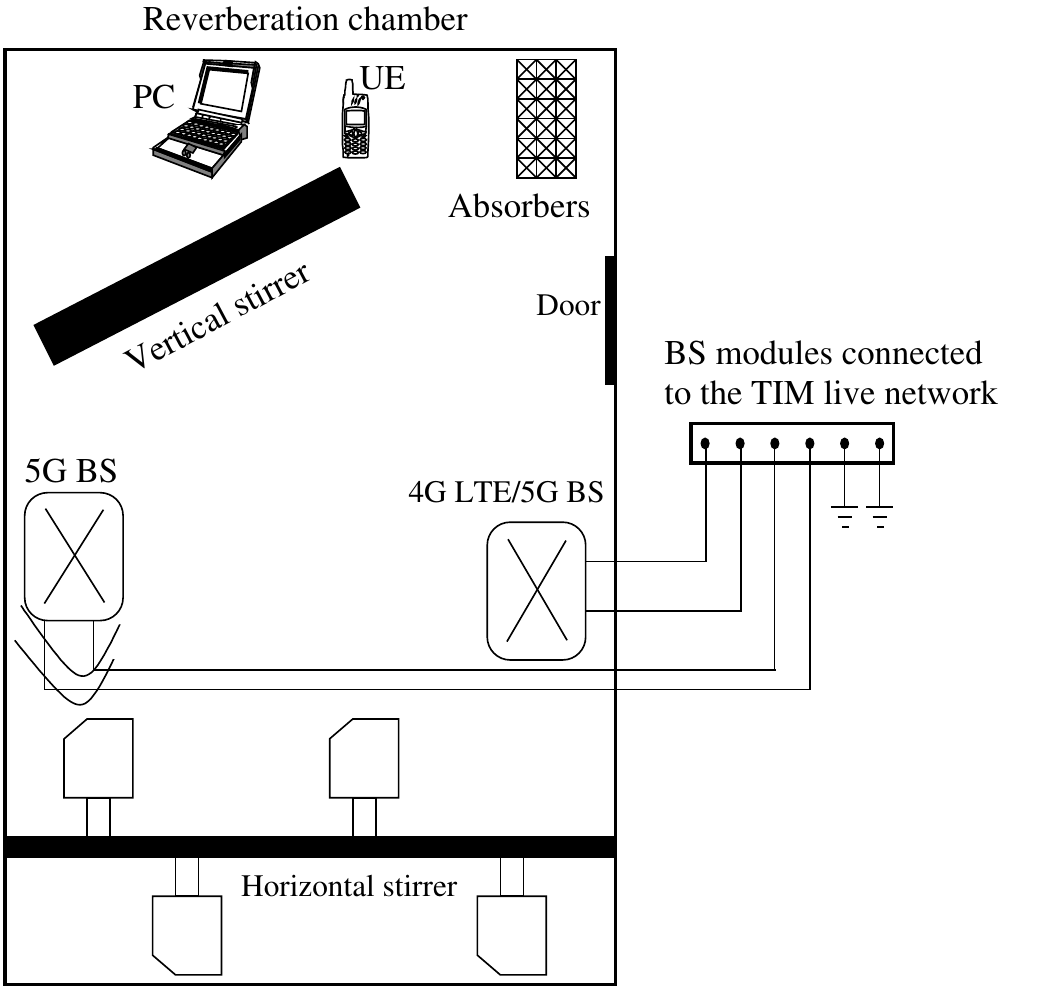}
\caption{Sketch of the experimental set-up. The notebook inside
the chamber is equipped by a USB data card and a software test tools connected to the EU.
The BSs are connected by an optic fiber to the Telecom Italia live network.}
\label{fig:schematico}
\end{figure}

\begin{figure}
\centering
\includegraphics[width=0.8\columnwidth]{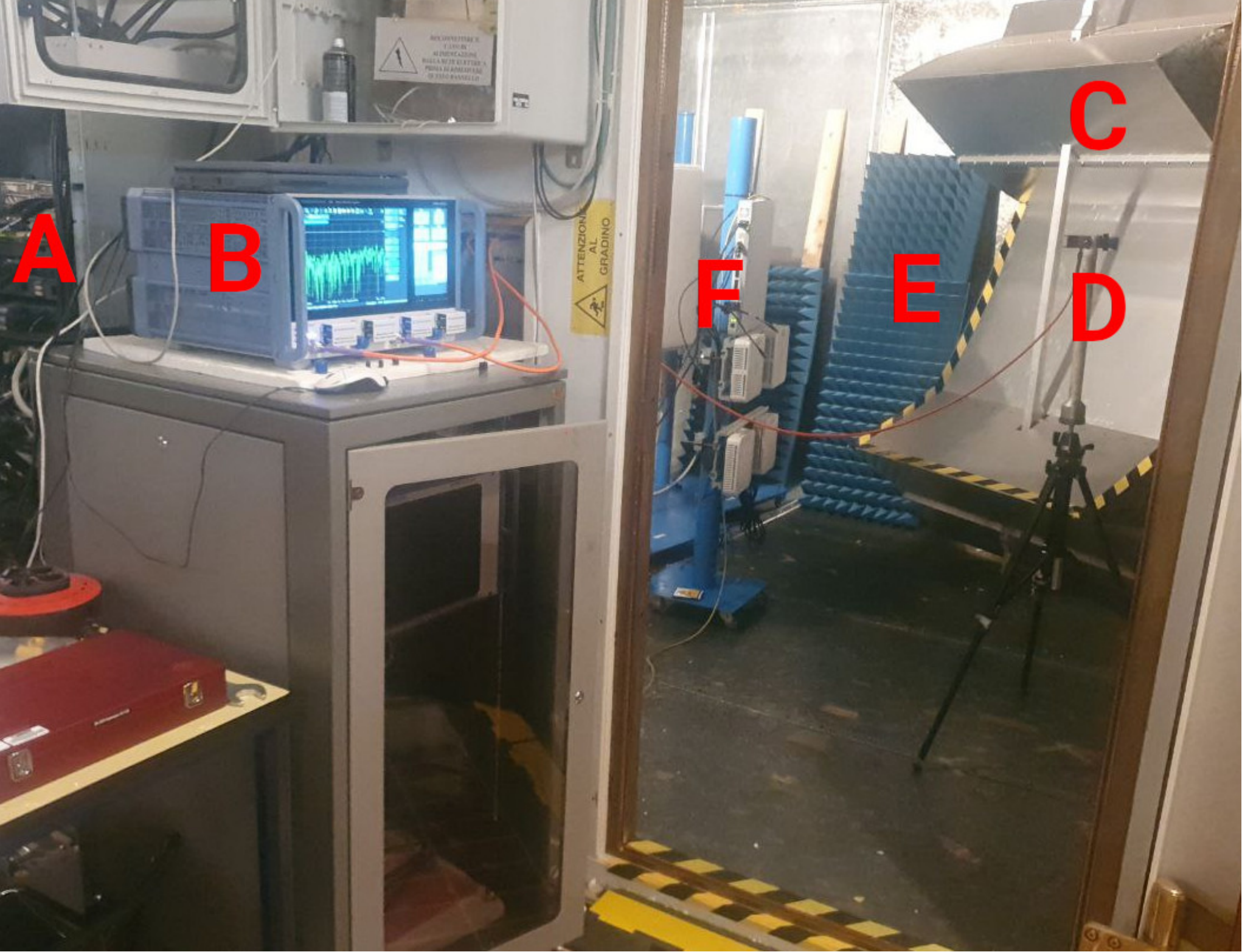}
\caption{Outer view picture of the set-up with the BSs rack (``A''), the VNA (``B''), the vertical stirrer (``C''), the horn antenna (``D''), pyramidal absorbers (''E'') and BSs (``F'').}
\label{fig:foto}
\end{figure}

\begin{figure}
\centering
\includegraphics[width=0.8\columnwidth]{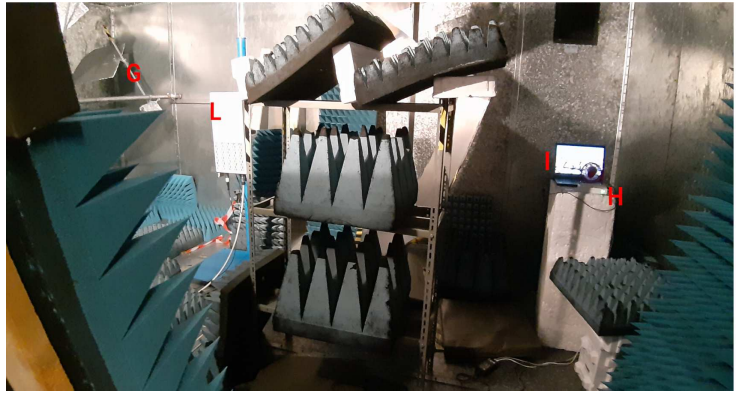}
\caption{Inner view of the RC loaded by absorbing materials and by the horizontal stirrer (``G''), the PC (``I''), the UE (``H'') and the 5G antenna (``L'').}
\label{fig:foto2}
\end{figure}

Figure~\ref{fig:schematico} reports the schematic block diagram of the measurements set-up.
The RC used for tests is placed at the EMC laboratory of Universit\`a Politecnica delle Marche and has dimensions of $6 \times 4 \times 2.5$~m$^3$. 
Within the RC are placed:
\begin{itemize}
\item 4G LTE and 5G base stations (BSs);
\item user equipment (UE);
\item PC;
\item pyramidal absorbers;
\item vertical and horizontal stirrers.
\end{itemize}
Figure~\ref{fig:foto} shows the external view of the RC whereas \figurename~\ref{fig:foto2} depicts the inner view of the chamber when it is loaded by absorbing materials.

During tests, the loading condition of the chamber were modified adding or removing electromagnetic absorbing materials, i.e. pyramidal absorbers.
In this way we are able to replicate a real-life environments within the RC.
In particular, we consider the empty chamber and a loaded chamber which emulates a residential environment.
In the second case we placed within the RC:
\begin{itemize}
\item $45$ anechoic absorbing panels (Emerson \& Cuming VHP-18-NRL);
\item $4$ anechoic absorbing panels (Emerson \& Cuming VHP-18-NRL);
\item $7$ planar absorbing panels (Emerson \& Cuming ANW-77).
\end{itemize}
The characterization of the RC in order to emulate the desired environment is done by adjusting the load and by matching the time delay spread $(\tau)$
according to international standards~\cite{guide1,guide2,guide3}.
Values of the $\tau_{RMS}$ (by considering a threshold of -20~dB~\cite{genender}) are reported in~Table~\ref{tab:tds}, sufficiently low to replicate a typical residential environment, according to standards on the investigated
frequency band~\cite{guide1,guide2,guide3}.

In order to evaluate the $(\tau)$ we collected the $S_{21}$ parameters by means of the vector network analyzer (VNA) and two horn antennas.
The antenna orientation were in non-line of sight (NLOS).
The chamber impulse response is evaluated by the inverse Fourier transform of the $S_{21}$ and then the power delay profile (PDP) is given by
\begin{equation} 
\textrm{PDP} \left ( t \right )=\left \langle \left | h \left ( t \right ) \right |^2 \right \rangle _N \,\, ,
\end{equation}
and the time delay spread root mean square $(\tau_{RMS})$ is evaluated by
\begin{equation}
\tau_{RMS} = \frac{\sqrt{\int_0^\infty {\left ( t - \tau_{ave} \right )^2 \textrm{PDP} \left ( t \right )dt}}} {\int_0^\infty { \textrm{PDP} \left ( t \right )dt}} \,\, ,
\end{equation}
where 
\begin{equation}
\tau_{ave} =\frac{ \int_0^\infty {t \textrm{PDP} \left (t\right ) dt}}{\int_0^\infty { \textrm{PDP} \left (t\right ) dt}} \,\, ,
\label{eq:tau_ave}
\end{equation}
the PDP is evaluated over the $N$ stirrer positions and $\left < \cdot \right >$ denotes the ensemble average over $N$.
\begin{table}
\caption{Measured time delay spread}
\label{tab:tds}
 \begin{tabular}{c|l|c}
  Frequency (GHz) & Environment &  Time delay spread {(ns)} \\ \hline
  28  & Empty chamber   &   138 \\ 
  28  & Loaded chamber  &   19  \\ 
 \end{tabular}
\end{table}
\begin{figure}
\centering
\includegraphics[width=0.8\columnwidth]{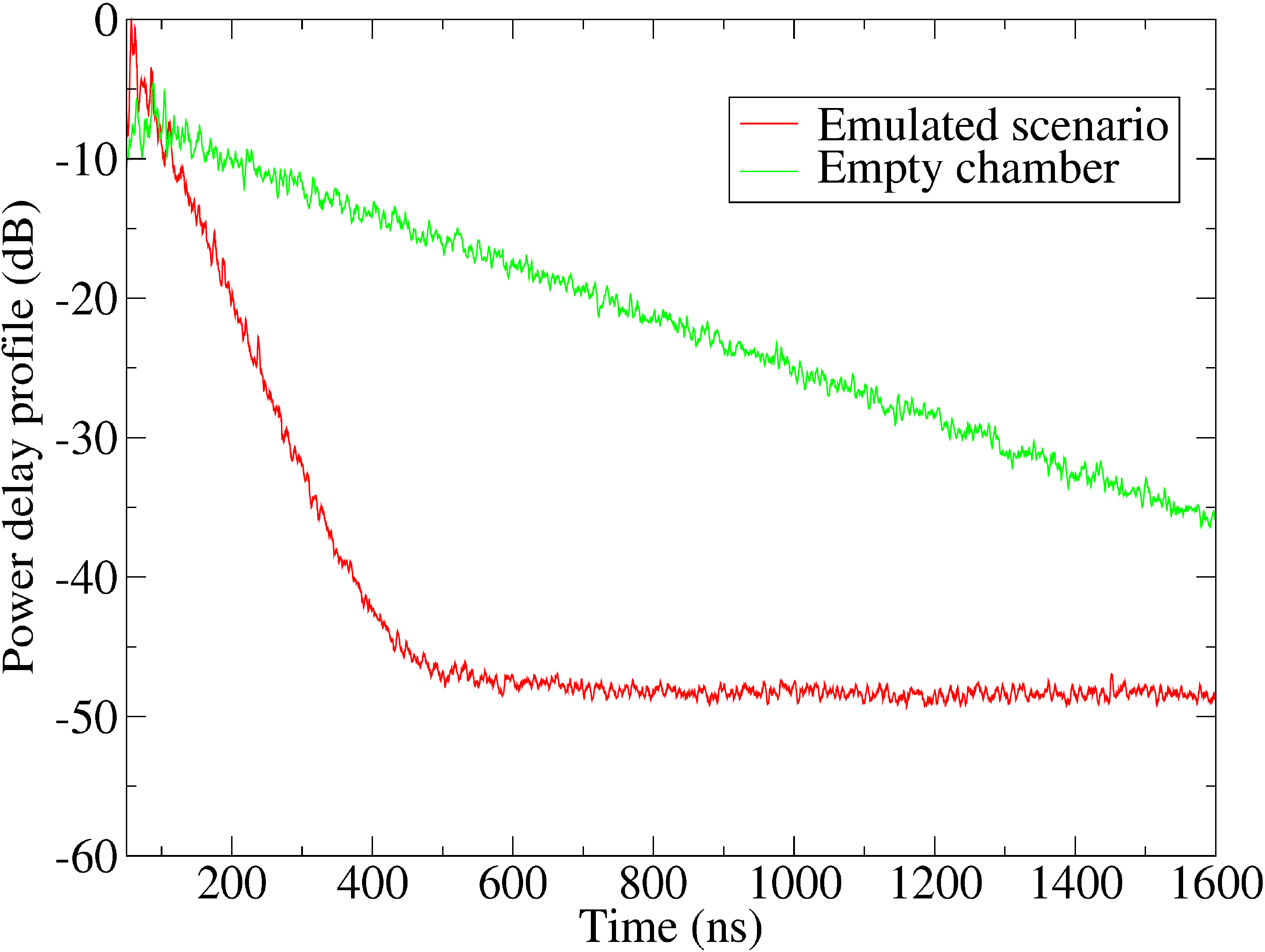}
\caption{Power delay profile of the empty chamber, green curve, and for the loaded chamber, red curve.}
\label{fig:pdp}
\end{figure}
The PDP of the empty and loaded chamber is reported in \figurename~\ref{fig:pdp}.
The addition of a large number of absorbing material considerably changes the PDP w.r.t. the empty chamber. 

\begin{figure}
\centering
\includegraphics[width=0.9\columnwidth]{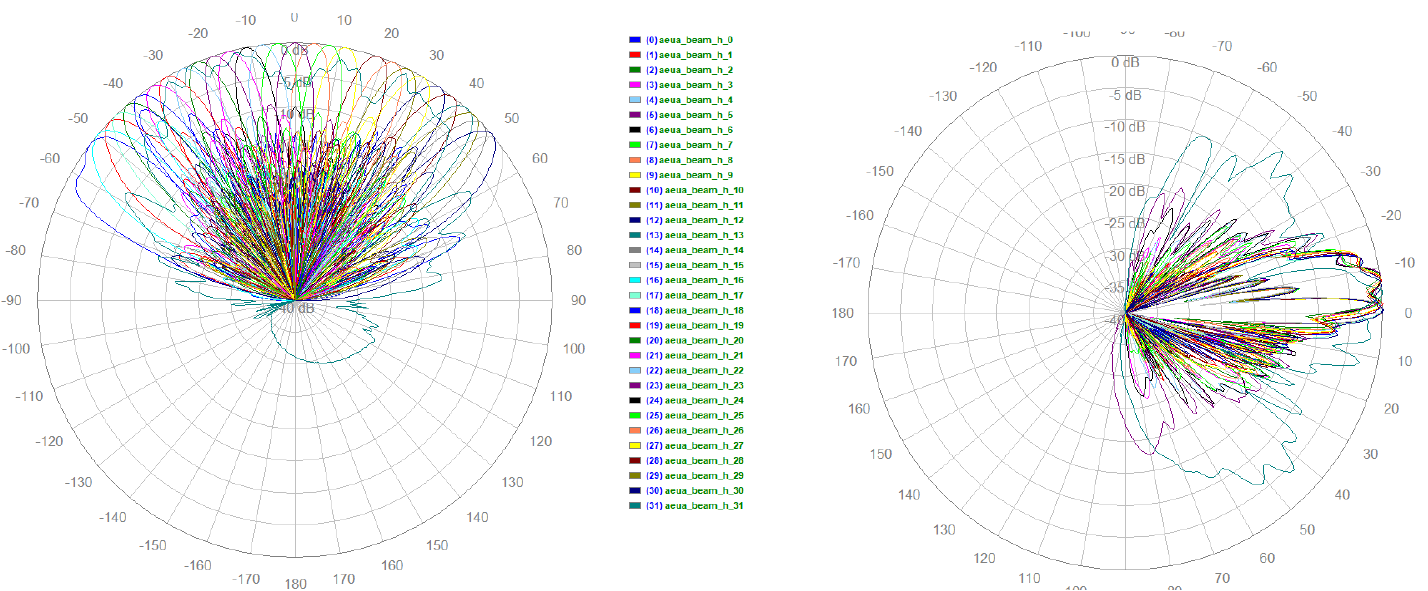}
\caption{Radiation pattern of the 32 beams, horizontal (left) and vertical (right) envelop beamwidth.}
\label{fig:beam}
\end{figure}

The RC is fed by the Nokia 5G BS and data were collected by the UE that is connected to the PC, see Figs.~\ref{fig:foto} and~\ref{fig:foto2}.
On the PC run a dedicated software that recorded the main parameters of the 5G signal.
The operation frequency of the BS is $26.95$~GHz.
The AEUB Nokia implements the analog beamforming, composed by $32$ beams with an azimuth opening angle from $-60$~deg to $+60$~deg and elevation opening angle from $+87$~deg to $+123$~deg.
Figure~\ref{fig:beam} reports the horizontal and vertical envelop of the 32 beams of the 5G antenna. 
On the horizontal plane, the antenna covers almost the full range.
The main preliminary testing was aimed to check the reaction capability of the 5G BS when the RC propagation channel is changed.
During measurements both horizontal and vertical stirrers have remained stationary.
When a residential scenario was emulated, the measured the RSRS and SNR were -59.1~dB and -10.8~dB respectively.
Subsequently, the horizontal stirrer was rotated in a continuous way with a speed of 10~deg/s.
Thanks to the capability of the BS, the system selects and switches to the optimal beam in order to maintain a good connection, according to the steps indicate in~\figurename~\ref{fig:beam}
In the same chamber configuration, during the horizontal stirrer rotation the average RSPR and SNR were -59.8~dB and -12.7 dB respectively.
The RSRP was not strictly affected by the dynamic multipath condition generated by the stirrer rotation whereas the SNR was reduced.
Results highlight the goodness of this BS feature in switching the most convenient antenna beam to looking for the best connection channel noised by a rich multipath.
The beam switching behavior has to be further investigated, a fortiori when the RIS will be integrated in the measurement set-up.
In future activities could be useful to fix the transmitting beam if we want to approximate a line of sight (LOS) scenario between the BS and the RIS instead of allowing the BS to continuosly evaluate and fix the ``best'' beam.

\section{Conclusion}

Performed tests were conducted in order to evaluate the performance of a 5G BS that operates at mm-wave frequencies.
Moreover, tests give us the possibility to investigate the analog beamforming functionalities of the BS. 
The adoption of the RC as facility test permits to emulate many real-life environments without moving the equipment every time thus saving time and money.
Preliminary results allow us to explore different situations with the goal of putting the RIS within the RC in order to evaluate its effect in a radio mobile communication in the upcoming activities. 

\section*{Acknowledgment}
This work has been supported by EU H2020 RISE-6G project under the grant number 101017011.


\end{document}